\documentstyle[12pt]{article}
\newcommand{\ka}{\kappa}
\newcommand{\la}{\lambda}
\newcommand{\de}{\delta}
\newcommand{\ta}{\tau}
\newcommand{\ze}{\zeta}
\newcommand{\La}{\Lambda}
\newcommand{\Th}{\Theta}
\newcommand{\om}{\omega}
\newcommand{\beq}{\begin{eqnarray}}
\newcommand{\eeq}{\end{eqnarray}}

\newcommand{\al}{\alpha}
\newcommand{\Ga}{\Gamma}
\newcommand{\ga}{\gamma}
\newcommand{\si}{\sigma}
\newcommand{\fhi}{\varphi}
\newcommand{\bet}{\beta}
\newcommand{\qq}{\mid q \mid}

\newlength{\extraspace}
\newlength{\extraspaces}
\newcounter{dummy}
\newcounter{cont}

\newcommand{\be}{\addtocounter{equation}{1}
\setcounter{cont}{\value{equation}}
\setcounter{equation}{0}
\renewcommand{\theequation}{\arabic{section}.\arabic{cont}}
\begin{equation}
\addtolength{\abovedisplayskip}{\extraspaces}
\addtolength{\belowdisplayskip}{\extraspaces}
\addtolength{\abovedisplayshortskip}{\extraspace}
\addtolength{\belowdisplayshortskip}{\extraspace}}
\newcommand{\ee}{\end{equation}\setcounter{equation}{\value{cont}}
\renewcommand{\theequation}{\arabic{section}.\arabic{equation}}}

\newcommand{\ai}{\addtocounter{equation}{1}
\setcounter{dummy}{\value{equation}}
\setcounter{equation}{0}
\renewcommand{\theequation}{\arabic{section}.\arabic{dummy}\alph{equation}}
\begin{eqnarray}
\addtolength{\abovedisplayskip}{\extraspaces}
\addtolength{\belowdisplayskip}{\extraspaces}
\addtolength{\abovedisplayshortskip}{\extraspace}
\addtolength{\belowdisplayshortskip}{\extraspace}}
\newcommand{\bj}{\end{eqnarray}\setcounter{equation}{\value{dummy}}
\renewcommand{\theequation}{\arabic{section}.\arabic{equation}}}

\begin{document}
\pagestyle{myheadings}

\title{Scalar-QED $\bet$-functions near Planck's Scale $^{a}$. \\ }

\author{Gentil O. Pires, \\
       {\small Departamento de Campos e Part\'\i culas (DCP),} \\
       {\small Centro Brasileiro de Pesquisas F\'\i sicas - CBPF/CNPq,} \\
       {\small Rua Dr. Xavier Sigaud 150, Urca,} \\
       {\small 22290-180 Rio de Janeiro - RJ, Brazil.} \\
       {\small \tt gentil@cat.cbpf.br}
       }

\footnotetext[1]{This work was completed at the Instituto de F\'\i sica,
                 Universidade Federal do Rio de Janeiro, Brazil.}

\date{ }

\maketitle

\begin{abstract}
\noindent
{\it The Renormalization Group Flow Equations of the Scalar-QED model near
     Planck's scale are computed within the framework of the
     average effective action. Exact Flow Equations, corrected by
     Einstein Gravity, for the running self-interacting
     scalar coupling parameter and for the running v.e.v. of $\phi^* \phi$,
     are computed taking into account threshold effects.
     Analytic solutions are given in the infrared and ultraviolet limits.
 }
\end{abstract}

\noindent {\small Pacs: 4.60.-m; 11.10.Hi; 11.25.Sq}

\section{ Introduction.}

\noindent
Nowadays, the satisfactory state of comprehension of the fundamental
interactions of nature based on the gauge principle constitutes a strong
appeal to quantize the gravitational field. Also, the search for a 
grand-unified theory in four dimensions shall demonstrate a complete 
sense only
when the quantization program for the gravitational field has reached
the same status of comprehension and consistency of the other three 
interactions. During the `70s, attempts to verify the renormalizability
of Einstein gravity as a perturbative quantum field theory were able to 
show the casual on-shell finiteness of pure gravity at 
one-loop \cite{t'Hooft},
the explicit loss of renormalizability at two-loops \cite{Sagnotti}, by the
generation of non-trivial higher-order counterterms that can not be 
absorbed in a redefinition of the physical parameters, and that 
matter-gravity coupled theories do not renormalize at all \cite{t'Hooft, Deser}.
In an attempt to cure the non-renormalizability of Einstein gravity
in four dimensions, some higher-derivative models were proposed 
\cite{Stelle, Shapiro}; these models were shown to be non-unitary,
though renormalizable, or unitary but now non-renormalizable again 
\cite{Nieuwenhuizen}.

Within this four dimensional picture, of apparent incompatibility of
quantum mechanics and gravity,  and having in mind that experiments
can only probe a limited range of energies, we are led to take a more
realistic position and interpret general relativity as an effective 
theory  \cite{Donoghue} up to Planck's scale, where low-energy scales 
are separated from unknown high-energy physics. The low-energy effective
theory, obtained integrating out unknow high-energy physics, 
does not present 
problems with renormalizability and/or unitarity. 
This can be easily 
understood in the following way : a theory of gravity
where high-energy effects are taken into account, presents high-derivative
sectors. These are responsible for the renormalizability of the model,
if one makes efforts to constrain its coefficients in this direction,
but unavoidably show the existence of massive ghosts.
As these ghosts have masses proportional to the Planck's mass, they shall
not be excited in the low-energy regime \cite{Percacci}. At sufficiently
low energies, these high-derivative corrections are not even relevant
to Physics phenomena at this lower scale \cite{Donoghue}.

Owing to the description of physics in a grand-unified picture, the effects
due to the gravitational interaction on field theory models become 
relevant when the energy scale is close enough to Planck's scale. In this 
scenario of taking into account gravity corrections at the grand-unified
range and the need for consistency in a quantum field theory description
of general relativity, the viewpoint we shall adopt is that of Wilson's
conception of an effective theory \cite{Wilson}, where the 
physical phenomena
should  be analyzed at a characteristic scale where its effects can 
effectively be verified (in contrast to the ``fundamental'' thought of
taking all scales at a time that has become the orthodoxy in quantum 
field theory) and the relation between different scales  described
by  ``exact renormalization group equations'' \cite{Polchinski}.

Recently, a new effective field theory approach in continuous space
\cite{Wetterich}, improved from Polchinski's presentation of the 
``flow equations'' \cite{Polchinski}, in the sense it gives the 
generating functional of
1PI-Green's functions as one runs the characteristic scale $\ka$
to zero, was proposed to exhibit infrared properties of theories such as
the convexity of the effective action when spontaneous symmetry breaking
takes place \cite{Wetterich2}.  This is known by the name of 
``average effective action''.

The average effective action $ \Ga_{\ka} $  takes averages of fields
in the effective action in continuous space 
\cite{Wetterich}-\cite{Wetterich2} by adding an infrared
smooth cutoff $ \Delta S_{\ka} $ to the action $ S $; so that all 
contributions to the effective action with momenta $ q^{2} < \ka^{2} $
are effectively suppressed. In the limit $ \kappa \rightarrow 0 $, 
the cutoff vanishes. $ \Delta S_{\ka} $ is written as 

\be
\label{cutoff}
\Delta S_{\ka} = \frac{1}{2} \int d^{4}\!q \; \phi_{_{A}} (-q) \;
R^{^{ [ A B ] }}_{\ka} (q) \; \phi_{_{B}} (q) \, ,
\ee

\noindent 
where $ \phi_{_{ A}} $ are generic fields whose Lorentz character
is described by the labels $ A $ and $ B $. $ R_{\ka} $ is defined 
in such a way as to imply that the 1PI-effective action is recovered from 
$ \Ga_{\ka} $ when the limit  $ \kappa \rightarrow 0 $ is taken,
independently of the form of  $ R_{\ka} $, and 
$ \Ga_{_{\ka \rightarrow \infty } } [ \phi ] = S [ \phi ] $ 
in the opposite limit \cite{Wetterich2}; showing that a flow equation can be 
written with solutions interpolating from the classical action to the 
effective action. Introducing sources to the action 
$ S + \Delta S_{\ka} $,
and defining the generating functional of 1PI-Green's 
functions by a Legendre
transformation of the $ \ka $-dependent generating functional of
connected Green's functions, one easily gets the exact evolution equations,
or flow equations, for the average effective action \cite{Wetterich2}

\be
\label{Flow}
\ka \frac{\partial}{\partial \ka} \Ga_{\ka} [ \phi ] 
= \frac{1}{2} Tr \; \left[ \left( \Ga_{\ka}^{(2)} [ \phi ] + 
R_{\ka} \right)^{-1} \; \ka  
\frac{\partial}{\partial \ka} R_{\ka} \right] \;  ,
\ee

\noindent
that relates phenomena at different scales. $ \Ga_{\ka}^{(2)} $ 
is the second
functional derivative of the average effective action with 
respect to the 
classical field $ \phi $; it is interpreted as the inverse 
full propagator.

In order to avoid some possible problems discussed in  
\cite{Percacci},
that may occur when explicitly stating  $  R_{\ka} $, we are 
going to use here a further
consideration: the squared  momentum contributions, $ q^{2} $, 
that we find in all 
inverse propagator coefficients obtained from the linearized 
form of the action 
$ S + \Delta S_{\ka} $, shall be replaced by the function

\be
P_{\ka} (q^{2})  = \frac{ q^{2} }{ 1 - f_{\ka}^{2}(q) } \; ,
\ee

\noindent  where

\be
 f_{\ka}(q) = exp \left[ - a \left( \frac{ q^{2} }{ \ka^{2}} \right)^{b} 
\right ] \; ,
\ee

\noindent with $ a $ and $ b $ constants, as the only effect that the 
cutoff term $  \Delta S_{\ka} $
has over the effective action \cite{Percacci}.

For $ q^{2} > \ka^{2} $, $ P_{\ka}(q^{2}) $ goes exponentially 
fast to $ q^{2} $, while 
for $ q^{2} < \ka^{2} $ the low-energy modes are suppressed. 
The function $ f_{\ka}(q) $
behaves like a modulating function acting on $ q^{2} $ varying 
from a Gaussian function,
when $ a = \frac{1}{2} $ and $ b = 1 $, for example, where 
$ P_{\ka}(q^{2}) $ tends to $ \ka^{2} $
for $ q^{2} \rightarrow 0 $, to a step function, when 
$ b \rightarrow \infty $, forbidding the modes with 
$ q^{2} < \ka^{2} $ from propagation. So, thanks to this 
behavior and  the $\ka $-derivative  of 
$ P_{\ka}(q^{2}) $ (see below), that suppress the 
propagation of large momentum fluctuations,
the momentum integration of the flow equations is 
ultraviolet and infrared finite. Thus,
no regularization scheme is needed in order to deal 
with high-energy quantum fluctuations;
provided an appropriate non-trivial cutoff term modulate 
the asymptotic behavior of effective
propagators and vertices  in such a way that the loop 
integrals of the flow equations are performed, effectively,
over a finite number of degrees of freedom. In this 
sense, the method is close to what is 
called non-perturbative field theory.

The average effective action, as proposed above, might 
turn the exact solution 
of the flow equations impossible to be found out due  
to an infinite number of effective insertion
terms that can be added to the effective action. Using 
some symmetry requirements along with 
the understanding that experiments go over a limited 
range of energies, one is then led to
consider only a finite and small number of relevant 
parameters that characterize physical 
phenomena up to a given energy scale  by performing 
a truncation scheme at a given scale
$ \ka_{1} $. The integration of the flow  equations 
of the effective action down
to $ \ka_{2} < \ka_{1} $ is, in principle, possible and 
so is the description of the flow of
the relevant parameters. 

In this work, we follow the approach of \cite{Reuter} 
when describing gauge fields by
the background field method, where it is shown that the 
gauge symmetry is preserved with 
respect to the gauge transformation of the background.

The paper is planned as follows :
in section 2, we define the average effective potential
at one-loop and extract the flow of relevant 
parameters in a general form. In the 3$^{rd.}$ and
4$^{th.}$ sections, we compute the flow equations for the 
Scalar-QED model with and without gravity interaction.
In section 5, we present our concluding remarks.

\section{Flow Equations.}

\noindent
Given the general form of the exact evolution equations 
(\ref{Flow}) above,
we can now introduce a set of prescriptions by imposing 
some definitions, or parametrizations,
for the first few parameters appearing in the  
average effective action $ \Ga_{\ka} $.
Remembering that we want to describe the flow  of the relevant 
parameters in a
spontaneously broken regime, let us parametrize the action by the 
minimum of its potential and the quartic self-interacting
coupling at the minimum.
In this phase, the average effective potential $ V_{\ka} $ has 
its minimum at $ \rho_{\ka} = (\fhi^{*} \fhi)_{\ka} $.
So,

\beq
\setcounter{equation}{1}
\label{Parametrizations}
V'_{\ka}(\rho_{\ka}) &  =  & 0 \; \; \;   , \;\;  
\la_{\ka} = V''_{\ka}(\rho_{\ka}) \; \; \;  ;
\\ \nonumber
\eeq

\noindent where each prime denotes a derivative with respect 
to $ \rho $.

The flow equations for $ \rho_{\ka} $ and $ \la_{\ka} $
can be read off by taking derivatives of  
(\ref{Parametrizations}) with respect
to $ \ka $ as below:

\beq
\label{Beta1}
\ka \frac{\partial}{\partial \ka} \rho_{\ka}
& = &
- \frac{1}{\la_{\ka}} \left( \ka \frac{\partial}{\partial \ka} 
V'_{\ka} \right)_{\rho \ = \ \rho_{\ka} }  
\;  \equiv \; 
\ka^{2}   \ga (\ka) \; \; , 
\eeq

\beq 
\label{Beta2}
\ka \frac{\partial}{\partial \ka} \la_{\ka}
& = &
\left( \ka \frac{\partial}{\partial \ka} 
V''_{\ka} \right)_{\rho \ = \ \rho_{\ka} }  
\;  \equiv \;  \bet (\ka) \; \; .
\eeq

\noindent 
In the definition of  $ \bet (\ka) $ we neglected a third derivative 
term which accounts for the variation
in the point of definition of $ \la_{\ka} $; we consider it an 
irrelevant contribution to the effective action.

As it can be seen, the evolution equation of $ \rho_{\ka} $ 
and $ \la_{\ka} $
can be computed using the definitions
above directly from the average effective potential $ V_{\ka} $. This
potential shall be given only at one-loop order, 
although it 
can be represented by one-loop Feynman diagrams of full propagators
\cite{Wetterich2}. To this end, we consider the effective action at a scale
$ \ka $ as the classical action  at a lower scale \cite{Percacci}.
Using the background field method, we write down the linearized form
of the Euclidean classical action, expand the effective action in powers
of momentum around the position of vanishing external momenta,
choose a specific configuration of the fields, so that only translationally
invariant vacuum expectation values are taken into account 
\cite{Coleman}, and properly 
define the determinant of the operators of small fluctuations
(see below) as to account 
for the volume of the spacetime $ \Omega $.
 
The linearized quadratic action, including Faddeev-Popov ghosts, 
gauge-fixing sectors and potential terms, can be written as

\be
\label{Quadraticaction}
S^{(2)} = \int d^{4}\!q \; \phi_{_{ A}} (-q) \;
a^{^{  A B }}_{i j} (J^{ P}) \; 
P^{^{  A B }}_{i j} (J^{ P}) \; \phi_{_{ B}} (q) \, ,
\ee

\noindent 
where $ a^{^{  A B }}_{i j} (J^{ P}) $ are coefficient matrices
that represent inverse propagators with definite spin and parity; 
$ P^{^{  A B }}_{i j} (J^{ P}) $ are spin-projection operators
\cite{Nieuwenhuizen, Percacci, Barnes, ECS}
listed in the appendix and the indices $ A, B $ label field
fluctuations above the background.

The  one-loop average effective action is obtained through a Legendre 
transformation of the generating functional of connected Green's functions
and an appropriate choice of $ \Delta S_{\ka} $ so
as to make possible the substitution of $ q^{2} $ by $ P_{\ka} ( q^{2} ) $
in the operator of small fluctuations  $ {\cal O}_{_{ A B}} $ 
defined below:

\be 
S^{(2)} = \int d^{4}\!q \; \sum_{ A, B } \; 
\phi_{_{A}} (-q) \;
{\cal O}_{_{ A B }}  \; 
\phi_{_{B}} (q) \, .
\ee

\noindent
So, provided an ultraviolet cutoff $ \La $ is taken, we define the
one-loop average effective action as

\be
\Ga_{\ka} ( \rho ) = \frac{1}{2} \ln \left( 
\frac{\det_{\ka} {\cal O} ( \rho ) }{ \det_{\ka} {\cal O} 
( \rho_{\ka} ) } \right) \; ,
\ee

\noindent
where $ \rho_{\ka} $ is the minimum of $ V_{\ka} $ and 
$ \det_{\ka} $ is the determinant with momentum integrations
modified as described above. With this definition for
$ \Ga_{\ka} $, with the normalization denominator
$ \det_{\ka} {\cal O} ( \rho_{\ka} ) $
included, the minimum of the
average effective potential is zero for all scales,

\be
V_{\ka} ( \rho_{\ka} ) = 0 \; ;
\ee

\noindent
this is consistent with the set of parametrizations adopted in 
(\ref{Parametrizations}).

Now, the average effective potential at one-loop can be written as

\be
\label{Potential}
V_{\ka} ( \rho ) = \frac{1}{2} \sum_{ J, P} \, 
\left( 2J + 1 \right) \,
\int_{_{ \mid q^{2} \mid = 0 }}^{^{ \La }} 
\frac{ d^{4}\!q }{ ( 2 \pi )^{4} } \, 
\ln \left( 
\frac{\det_{\ka} a_{\ka} ( J^{P} ) ( \rho )}{\det_{\ka} a_{\ka} 
( J^{P} ) ( \rho_{\ka} )} 
\right) \; ,
\ee

\noindent
where  $ ( 2J + 1 ) $ stands for all the multiplicity of each spin
contribution.

Given the definition for $ V_{\ka}(\rho) $ 
as above and the inverse  propagator coefficients,
$ a_{\ka}(J^{P}) $, stemming from the linear
quadratic action (\ref{Quadraticaction}), we are able to
compute the $ \bet $-functions (\ref{Beta1}) and (\ref{Beta2})  and 
analyze the scale-dependence of the parameters that characterize the
effective action. The general functions are

\beq
\label{Flow2}
\ga (\ka) & = & \frac{- 1}{ 32 \pi^{2} \, \ka^{2} \, \la_{\ka}} \, 
\int \, dx \, x {\cal R}_{\ga} (P_{\ka}, \rho_{\ka}) \, 
\ka \frac{\partial P_{\ka}}{ \partial \ka }\;  , 
\eeq

\beq
\label{Flow3}
\bet(\ka) & = & \frac{1}{ 32 \pi^{2} } \, 
\int \, dx \, x {\cal R}_{\bet} (P_{\ka}, \rho_{\ka}) \, 
\ka \frac{\partial P_{\ka}}{ \partial \ka } \;  ;
\eeq

\noindent
where $ x = \mid q^{2} \mid  $ and $ {\cal R } $ are 
rational functions of $ P_{\ka} $ and $ \rho_{\ka} $. 
Notice that as a by-product of our 
appropriate infrared cutoff $ \Delta S_{\ka} $ introduced at a 
scale $ \ka $, due to the behavior of $ P_{\ka} (x) $ and its 
k-derivative, the momentum integration in the flow equations
(\ref{Flow2}) and (\ref{Flow3}) are ultraviolet and infrared finite.
$ \ka \, \frac{\partial P_{\ka}}{\partial{\ka}} $
receives an effective contribution at $ x \approx \ka^{2} $
and large momentum fluctuations $ q^{2} \gg \ka^{2} $
are exponentially suppressed, while the $ \ka $-scale acts like
a mass term in the inverse propagators $ P_{\ka} (q^{2}) $ in 
the infrared limit $ q^{2} \ll  \ka^{2} $. Thus, only a finite
number of degrees of freedom effectively contributes to
(\ref{Flow2}) and (\ref{Flow3}), as an exact evolution equation should be.

\section{Scalar-QED.}

Let us now compute the flow equations, described in the previous
section, for the case of Scalar-QED. The effective
action we shall work with is given by

\beq
\setcounter{equation}{1}
\label{Classicalaction}
\Ga (A_{\mu}, \fhi^{*}, \fhi) 
& = & \nonumber
\int \, d^{4} x \; \left[ \; V (\rho) + \right.
Z_{\rho} (D_{\mu} \fhi)^{*} (D^{\mu} \fhi) +  \\
&   & \nonumber \\
& + & \left.
\frac{1}{4} Z_{F} F_{\mu \nu} F^{\mu \nu} +
\frac{1}{2 \alpha} (\partial_{\mu} A^{\mu})^{2} \; \right] \; ,
\eeq

\noindent 
where $ D_{\mu} \equiv \partial_{\mu} + i e_{o} A_{\mu} $
and  $ \rho \equiv \fhi^{*} \fhi $. $ Z_{\rho} $ and 
$ Z_{F} $ are fixed wave-function renormalization 
constants $^{b}$
\footnotetext[2]{ In this work, we do not compute the
running of the wave function renormalization constants $ Z_{F} $
and $ Z_{\rho} $. It was shown in ref. \cite{Percacci} that 
for the pure scalar field case, in the ultraviolet and infrared limits,
the anomalous dimension in four dimensions 
can be neglected; $ \eta \ll 1 $.
We set here $ Z_{\rho} = Z_{F} = 1 $ because we
don't expect it to give significant different results
in these limits.}
and $ e_{o} $ is a bare coupling constant. 

Following the 
steps described in the previous section, the associated
average effective potential is

\be
\label{Potential2}
V_{\ka} (\rho)  =  \frac{1}{2} \int \, 
\frac{d^{4} q}{(2 \pi)^{4}} \;
\left[ 3 \ln \left( 
\frac{ a_{\ka} ( 1^{-} ) ( \rho )}{ a_{\ka} 
( 1^{-} ) ( \rho_{\ka} )} \right)  + 
\ln \left( \frac{\det_{\ka} a_{\ka} ( 0^{+} ) 
( \rho )}{\det_{\ka} a_{\ka} 
( 0^{+} ) ( \rho_{\ka} ) } \right) \right] \; .
\ee

\noindent
Here, we have split the fields $ A_{\mu}, \fhi^{*} $ and $ \fhi $
in classical backgrounds, that will be identified as vacuum 
expectation values, plus fluctuations as below:

\beq
\label{Background}
\fhi^{*} & = & \fhi^{*}_{_{cl.}} + \de \fhi^{*} \; , \nonumber \\
\fhi & = & \fhi_{_{cl.}} + \de \fhi \; , \\
A_{\mu} & = & \de A_{\mu} \; ; \;  A_{\mu}^{^{cl.}} \equiv 0 \; .
\nonumber
\eeq

\noindent
The inverse average propagator coefficients $ a_{\ka} ( J^{P} ) ( \rho ) $
of eq. (\ref{Potential2}) are given by

\be 
\label{Coefficient1}
a_{\ka} ( 1^{-} ) = \sqrt{3} \left( P_{\ka} (q^{2}) + 
2 e_{o}^{2} \, \rho \right) \; ,
\ee

\be
\label{Coefficient2}
a_{\ka} ( 0^{+} ) = 
\left( \begin{array}{ccc}
\frac{1}{\al} P_{\ka} (q^{2}) + 2 e_{o}^{2} \rho & 
e_{o} \fhi^{*} \mid P_{\ka}^{\frac{1}{2}} (q^{2}) \mid  & 
- e_{o} \fhi \mid P_{\ka}^{\frac{1}{2}} (q^{2}) \mid \\
e_{o} \fhi^{*} \mid P_{\ka}^{\frac{1}{2}} (q^{2}) \mid   & 
(\fhi^{*})^{2} V'' (\rho)  & 
P_{\ka} (q^{2}) + V' (\rho) + \rho V'' (\rho) \\
- e_{o} \fhi \mid P_{\ka}^{\frac{1}{2}} (q^{2}) \mid & 
P_{\ka} (q^{2}) + V' (\rho) + \rho V'' (\rho) &
(\fhi)^{2} V'' (\rho)
\end{array}
\right) \; . 
\ee

\noindent 
They can be easily read off from the linearized quadratic action
in coordinate space

\beq
S^{(2)} (A_{\mu}; \fhi^{*}, \fhi ) 
& = &
\int \, d^{4} x \, \left[ \de A_{\mu} \left( ( - \partial^{2} )
\Th^{\mu \nu} + \frac{1}{\al} ( - \partial^{2} ) \, \om^{\mu \nu} +
2 e_{o}^{2} \, \rho_{_{cl.}} \, \de^{\mu \nu} \right) \, 
\de A_{\nu} \right. +
\nonumber \\ & + & 
2 \, \de \fhi^{*} \, \left( \, (-D^{2}) + V' (\rho) \mid_{_{\fhi_{cl.}}} + 
\rho_{_{cl.}} \, V'' (\rho) \mid_{_{\fhi_{cl.}}} \, \right) \, \de \fhi + 
\nonumber \\ & + &
V'' (\rho) \mid_{_{\fhi_{cl.}}} \, \left( (\de \fhi^{*})^{2} 
\fhi_{_{cl.}}^{2} +
(\fhi_{_{cl.}}^{*})^{2} (\de \fhi)^{2}  \right) +
\\ & + &
4 e_{o}^{2} \, A^{\mu}_{_{cl.}} \, \de A_{\mu} \,  
\left( \, \fhi_{_{cl.}} \, \de \fhi^{*}  + 
\fhi_{_{cl.}}^{*} \, \de \fhi \, \right) +
\nonumber \\ & - & 
\left.
2 \, i \, e_{o} \, \de A_{\mu} \, 
\left( \, \de \fhi^{*} \partial^{\mu} \fhi \mid_{_{\fhi_{cl.}}} - \,
\fhi_{_{cl.}} \partial^{\mu} \de \fhi^{*} + 
\fhi_{_{cl.}}^{*} \partial^{\mu} \de \fhi -
\de \fhi \, \partial^{\mu} \fhi^{*} 
\mid_{_{\fhi_{cl.}}} \, \right) \, \right] \, ;
\nonumber
\eeq

\noindent
after it is Fourier transformed to momentum space,
the different spin-parity
contributions$^{\, c}$
\footnotetext[3]{\noindent See eq.
(\ref{Quadraticaction}) and Appendix.}
are identified, the field configurations (\ref{Background})
are chosen and $ q^{2} $  is replaced by
$ P_{\ka} (q^{2}) $.

We call the readers attention to the fact that we have not
constrained any longitudinal mode in order to obtain 
manifest gauge invariant results. Terms containing
$ (\de \fhi)^{2} $, $ (\de \fhi^{*})^{2} $ and longitudinal
vector fields do propagate in our approach due to the presence
of an infrared cutoff. In this way, the gauge invariance could
be achieved as the limit $ \ka \rightarrow 0 $ of the
modified Ward-identities as in refs. \cite{Becchi}.

After taking derivatives of the coefficients (\ref{Coefficient1}) and
(\ref{Coefficient2}), with respect to $ \ka $ and $ \rho $ 
as in  eqs. (\ref{Beta1}) and 
(\ref{Beta2}), we substitute into the flow equations the `classical'
parameters  and the `classical' potential $ V $, from 
(\ref{Classicalaction}), by their running counterparts. In connection
to what was called a `classical' action, as an effective action
at a lower scale, the procedure can be iterated at each scale $ \ka $.
This updating is called the renormalization group improvement,
or, sometimes, fine-tunning, and the flow of $ V'_{\ka} (\rho) $  and
$ V''_{\ka} (\rho) $ with $ \ka $ shall be the flow equations
we are searching for.

Analytic solutions of eqs. (\ref{Flow2}) and (\ref{Flow3}) 
for this model in closed
form are not possible to be found. 
However, we are able to analyze asymptotic limits and obtain
analytic solutions for $ \ka^{2} $ very large or very small
compared to $ \rho_{\ka} $ . 

In the ultraviolet limit, where $ q^{2} \gg \rho_{\ka} $, the 
$ \ga $- and $ \bet $-functions describing the running of 
the v.e.v. of $ \rho_{\ka} $ and the running of
$ \la_{\ka} $ are dominated by powers of $ P_{\ka} (x) $,
which is of order $ \ka^{2} $,  since the factor 
$ \ka \frac{\partial \, P_{\ka}}{\partial \, \ka} $,
in eqs. (\ref{Flow2}) and (\ref{Flow3}),
suppressing exponentially large momentum fluctuations in
comparison to $ \ka $ and as a power for lower $ q $,
receives its effective contribution at $ q \approx \ka $, 
where it is peaked.
So, in this limit we neglect $ \rho_{\ka} $ with respect to
$ \ka^{2} $. The leading contributions are

\beq
\label{sq-6a}
\ga (\ka) & = & \frac{-1}{32 \, \pi^{2} \, \ka^{2} \, \la_{\ka} } \,
\int \, dx \, x \; \ka \, 
\frac{\partial \, P_{\ka} (x) }{\partial \, \ka} \,
\left[ \, - 6 e_{o}^{2} - 4 \al e_{o}^{2} - 4 \la_{\ka} \, \right] \,
P_{\ka}^{\, -2} (x) \, , 
\\ \nonumber & &  \\
\label{sq-6b}
\ka \frac{\partial \, \rho_{\ka}}{\partial \ka} 
& = &
\frac{\ka^{2}}{16 \pi^{2} \la_{\ka}} \, \left[ \, 3 e_{o}^{2} 
+ 2 \al e_{o}^{2} + 2 \la_{\ka} \, \right] \, I_{-2}(0)
\eeq

\vspace{.5cm}

\noindent and

\beq
\label{sq-7a}
\bet (\ka) & = & \frac{1}{32 \, \pi^{2} } \,
\int \, dx \, x \; \ka \, 
\frac{\partial \, P_{\ka} (x) }{\partial \, \ka} \,
\left[ \, 24 e_{o}^{4} + 32 \al^{2} e_{o}^{4} 
+ 8 \al e_{o}^{2} \la_{\ka} + 20 \la^{2}_{\ka} \, \right] \,
P_{\ka}^{\, -3} (x) \, , 
\\ \nonumber & &  \\
\label{sq-7b}
\ka \frac{\partial \, \la_{\ka}}{\partial \ka} 
& = & 
\frac{1}{8 \pi^{2} } \, \left[ \, 6 e_{o}^{4} + 8 \al^{2} e_{o}^{4} 
+ 2 \al e_{o}^{2} \la_{\ka} + 5 \la^{2}_{\ka} \, \right] \, I_{-3}(0) \, ;
\eeq

\vspace{.5cm}

\noindent where the integrals $ I_{n}(w) $ are defined by 

\beq
 \ka^{2(n+3)} \, I_{n}(w)  = \int  & dx \, x &  ( \,
P_{\ka} (x) + w \, )^{n} \,  \ka 
\frac{\partial P_{\ka} (x)}{\partial \ka}  \, , 
\\ \nonumber & & \\
 I_{-2} (0)  =  \ka^{-2} \, \int & dx \, x &  P_{\ka}^{\, -2} (x) \, 
\ka \frac{\partial P_{\ka} (x)}{\partial \ka} \, 
= \frac{2}{(2a)^{\frac{1}{b}}} \, \Ga (1 + \frac{1}{b} ) 
\eeq

\noindent and

\be
I_{-3} (0)  =  \, \int \, dx \, x  \, P_{\ka}^{\, -3} (x) \, 
\ka \frac{\partial P_{\ka} (x)}{\partial \ka} \, 
= \, 1 \, .
\ee

\noindent 
From this we see that,
independently of the constants $ a $ and $ b $,
$ \rho_{\ka} $ runs quadratically in
the ultraviolet regime, as dimensional arguments  suggest perturbatively,
and $ \la_{\ka} $ scales logarithmically with sublogarithmic
corrections at high energies.

There is an apparent incompatibility between the running of 
$ \rho_{\ka} $ and the approximation $ \ka^{2} \gg \rho_{\ka} $.
Looking at eqs. (\ref{sq-6a}) - (\ref{sq-7b}), we see that
$ \rho_{\ka} $ runs in fact quadratically, which makes the approximation
questionable, but  $ \la_{\ka} $ runs logarithmically with some 
sublogarithmic corrections. Thus, the net result is that $ \rho_{\ka} $
grows slower than $ \ka^{2} $ what justifies the approximation in
this case.

In the opposite limit, $ \ka^{2} \ll \rho_{\ka} $, the infrared one,
the flow equations are dominated by powers of $ \rho_{\ka} $
greater than powers of $ P_{\ka} (x) \approx \ka^{2} $. 
The leading contributions
are those with higher powers of $ \rho_{\ka} $. So, the 
$ \ga $-function is given by

\beq
\label{sq-9a}
\ga (\ka) & = & \frac{-1}{32 \, \pi^{2} \, \ka^{2} \, \la_{\ka} } \,
\int \, dx \, x \; \ka \, 
\frac{\partial \, P_{\ka} (x) }{\partial \, \ka} \,
\left[ \, \frac{- 3}{2 \rho_{\ka}^{2} e_{o}^{2}} - 
\frac{\la_{\ka}}{2 P_{\ka}^{\, 2} (x)}
\, \right] \, , 
\\ \nonumber & &  \\
\label{sq-9b}
\ka \frac{\partial \, \rho_{\ka}}{\partial \ka} 
& = & 
\frac{1}{32 \pi^{2} \la_{\ka}} \, \left[ \,
\frac{3}{2 \rho_{\ka}^{2} e_{o}^{2}} \, I_{\, 0} (0) \,  \ka^{6} +  
\frac{\la_{\ka}}{2} \, I_{\, -2} (0) \, \ka^{2} 
 \, \right] \, .
\eeq

\vspace{.5cm}

Solving the above p.d.e. for $ \la_{0} \neq 0 $ and $ \rho_{0} \neq 0 $,
we get that the first term on the r.h.s. of eq. (\ref{sq-9b}) 
becomes the dominant one 
which damps the scaling of $ \rho_{\ka} $ by powers of
$ \frac{\ka^{2}}{\rho_{\ka}} $ and stops its running for 
$ \ka \rightarrow 0 $.
For small enough $ \ka , \rho_{\ka} $ runs as

\be
\label{sq-10}
\rho_{\ka} = \rho_{0} \, \left[ \, 1 +  \frac{1}{128 \pi^{2}} \, 
\frac{I_{0}(0)}{\la_{0} e_{o}^{2}} \, 
\frac{\ka^{6}}{\rho_{0}^{3}} \, \right] \, ,
\ee

\noindent where

\be
I_{0} (0) = \ka^{\, - 6} \, \int \, dx \, x \; \ka \, 
\frac{\partial \, P_{\ka} (x) }{\partial \, \ka} \, 
\ee
is a constant.

\vspace{1cm}

The $ \bet $-function  in this limit is given by

\beq
\label{sq-12a}
\bet (\ka) & = & \frac{1}{32 \, \pi^{2} } \,
\int \, dx \, x \; \ka \, 
\frac{\partial \, P_{\ka} (x) }{\partial \, \ka} \,
\left[ \, \frac{3}{\rho_{\ka}^{3} e_{o}^{2}} + 
\frac{\la_{\ka}^{2}}{2 P_{\ka}^{\, 3} (x)} \, \right] \,  ,
\\ \nonumber & &  \\
\label{sq-12b}
\ka \frac{\partial \, \la_{\ka}}{\partial \ka}
& = & 
\frac{1}{32 \pi^{2} } \, \left[ \, 
\frac{3}{\rho_{\ka}^{3} e_{o}^{2}} \, \ka^{6} \, I_{\, 0} (0) + 
\frac{\la_{\ka}^{2}}{2} I_{\, -3} (0)
\, \right] \,  .
\eeq

\noindent
The first term in the r.h.s. contributes with the 
same behavior seen in the
pure scalar case \cite{Percacci}, although it comes from a 
purely vector contribution (spin-1) : the running of $ \la_{\ka} $ 
is damped by powers of $ \frac{\ka^{2}}{\rho_{\ka}} $, 
going to zero in the limit $ \ka \rightarrow 0 $. The 
second term contributes logarithmically to the scaling of
$ \la_{\ka} $; it comes from a longitudinal vector contribution and
show the same net behavior found in the 
ultraviolet regime. Thus, when $ \ka \rightarrow 0 $, the damped 
term decouples but the theory still correlates  at long
distances.

\vspace{.5cm}

\section{Turning on the Gravity interaction.}

We are now going to use the techniques shown in the sections before
to compute and analyze the $ \bet $-function of the 
Scalar-QED model when 
gravitational effects are taken into account;
i.e., we are going to correct the  $ \bet $-functions of the last
section when the energy scale is close to, but below, the Planck's
scale. To do that, we take our effective action as

\beq
\label{G-1}
\setcounter{equation}{1}
\Ga (A_{\mu}, \fhi^{*}, \fhi; h_{\mu \nu}) 
& = & \nonumber
\int \, d^{4} x \; \sqrt{g} \, \left[ \; V (\rho) + \right.
g^{\mu \nu} \, (D_{\mu} \fhi)^{*} (D_{\nu} \fhi) +  \\
&   & \nonumber \\
& + & \left.
\frac{1}{4} g^{\mu \nu} g^{\ka \la} \, F_{\mu \ka} F_{\nu \la} +
\frac{1}{2 \alpha}  g^{\mu \nu} g^{\ka \la} \, (\partial_{\mu} A_{\nu})  
(\partial_{\ka} A_{\la}) 
+  {\cal K} \, R\; \right] \; ,
\eeq

\noindent where, as before, 
$ D_{\mu} \equiv \partial_{\mu} + i e_{o} A_{\mu} $
and  $ \rho \equiv \fhi^{*} \fhi $. $ {\cal K} = \frac{1}{16 \pi G}, 
g = det \, g_{\mu \nu}, 
R = g^{\mu \nu} R_{\ga \mu \; \nu}^{\;\;\;\; \ga} $ is the
scalar curvature and 
$  R_{\mu \nu \; \si}^{\;\;\;\; \rho} = 
\partial_{\mu} \, \Ga_{\nu \;\; \si}^{\;\; \rho} -
\partial_{\nu} \, \Ga_{\mu \;\; \si}^{\;\; \rho} + 
\Ga_{\mu \;\; \ga}^{\;\; \rho} \, \Ga_{\nu \;\; \si}^{\;\; \ga} -
\Ga_{\nu \;\; \ga}^{\;\; \rho} \, \Ga_{\mu \;\; \si}^{\;\; \ga} $
is the curvature tensor, where the connection is defined as usual; 
$ \Ga_{\nu \;\; \si}^{\;\; \rho} = \frac{1}{2} \, g^{\rho \ta}
\left( \partial_{\nu} g_{\ta \si} + \partial_{\si} g_{\nu \ta} -
\partial_{\ta} g_{\si \nu} \right). $

Hereafter, we shall recognize the metric field $ g_{\mu \nu} $ as 
a fluctuation around the Euclidean (flat) space geometry,

\be
g_{\mu \nu} (x) = \de_{\mu \nu} + \frac{1}{\sqrt{{\cal K}}} \, 
h_{\mu \nu} (x) \, ,
\ee

\noindent and take $  h_{\mu \nu} (x) $ as the field variable of 
the gravitational interaction. Also, we shall expand the fields
$ A_{\mu}, \fhi^{*} \; $ and $ \; \fhi $ as in (\ref{Background}).

The linearized quadratic action is then written as

\beq
\label{G-3}
S^{(2)} (A_{\mu}; \fhi^{*}, \fhi; h_{\mu \nu}) 
& = &
\int \, d^{4} x \, 
\left[ 
\de A_{\mu} 
\left(
\de^{\mu \nu} \, \de^{\ka \la}  ( - \partial^{2}
\Th^{\mu \nu} ) \, + 
\frac{1}{\al} \de^{\mu \nu} \, \de^{\ka \la} 
( - \partial^{2} \, \om_{\nu \ka} ) \, +
\right.
\right.
\nonumber \\ & + & 
\left.
2 e_{o}^{2} \, \rho_{_{cl.}} \, \de^{\mu \la} 
\right) 
\, \de A_{\la} \,  +
\nonumber \\ & + & 
2 \, \de \fhi^{*} \, 
\left( 
\, - \de^{\mu \nu} \, \partial_{\mu} \, \partial_{\nu}  + 
V' (\rho) \mid_{_{\rho_{cl.}}} \, + 
\rho_{_{cl.}} \, V'' (\rho) \mid_{_{\rho_{cl.}}} \, 
\right) 
\, \de \fhi \, + 
\nonumber \\ & + &
V'' (\rho) \mid_{_{\rho_{cl.}}}    
\left( \,
(\de \fhi^{*})^{2} \fhi_{_{cl.}}^{2} +
(\fhi_{_{cl.}}^{*})^{2} (\de \fhi)^{2} \, 
\right) \, +
\nonumber \\ & - &
i \, 2 \, e_{o} \, \de A_{\mu} \, \de^{\mu \nu} 
\left( 
\, \fhi^{*}_{_{cl.}} \partial_{\nu} \, \de \fhi -
\fhi_{_{cl.}} \partial_{\nu} \, \de \fhi^{*} \, 
\right) \, +
\nonumber \\ & + &
\frac{1}{\sqrt{{\cal K}}} \, h^{\mu \nu} \, 
\left( \,
\frac{1}{2} \, \de_{\mu \nu} \,  V' (\rho) \mid_{_{\rho_{cl.}}} \,
\fhi^{*}_{_{cl.}} \,
\right) 
\de \fhi \, +
\nonumber \\ & + &
\frac{1}{\sqrt{{\cal K}}} \, \de \fhi^{*} \, 
\left( \,
\frac{1}{2} \, \de_{\mu \nu} \,  V' (\rho) \mid_{_{\rho_{cl.}}} \,
\fhi_{_{cl.}} \,
\right) \, 
h^{\mu \nu} \, +
\nonumber \\ & + &
\frac{1}{{\cal K}} \, h^{\mu \nu} \, 
\left( \, \frac{-1}{4} \, \de_{\mu \ka} \de_{\nu \la} \, V (\rho_{cl.}) \, +
\frac{1}{8} \, \de_{\mu \nu} \de_{\ka \la} \, V (\rho_{cl.}) \,
\right) \,
h^{\ka \la} \, +
\nonumber \\ & + &
\frac{1}{2} \,   h_{\mu \nu} \, \left( \, \de^{\mu \nu} \, 
\partial^{\rho} \, \partial^{\si} \, - \, \frac{1}{2} \, \de^{\mu \nu} 
\de^{\rho \si} \, \partial^{2} \, - \de^{\nu \si} \, 
\partial^{\mu} \, \partial^{\rho} \, +
\right.
\nonumber \\ & + &
\left.
\frac{1}{2} \, \de^{\mu \rho} \, \de^{\nu \si} \, \partial^{2} \,
\right) \,
h_{\rho \si} \, +
\frac{1}{2 \, \xi} \, \, \partial_{\mu} \, h^{\mu \nu} \, 
\partial^{\si} \, h_{\si \nu} 
\left.
\, 
\right] \, ,
\eeq

\noindent where we have fixed the gauge symmetry

\be
\de h_{\mu \nu} (x) = \partial_{\mu} \ze_{\nu} (x) + 
\partial_{\nu} \ze_{\mu} (x) \; ,
\ee

\noindent with $ \ze_{\nu} (x) $ being an infinitesimal coordinate
transformation, by adding the gravitational gauge-fixing sector

\be
S_{_{G.F.}} = \frac{1}{2 \, \xi} \, \int \, d^{4} x \, 
\partial_{\mu} \, h^{\mu \nu} \, \partial^{\si} \, h_{\si \nu} \, .
\ee

\noindent
The Faddeev-Popov ghosts decouple from $ h_{\mu \nu} $ in this gauge
and shall not be considered in what follows.

The linearized quadratic action (\ref{G-3}) can be rewritten 
in compact form spanning all its elements on the corresponding 
spin-projection operators as in eq. (\ref{Quadraticaction}). The 
coefficient matrices $ a^{A B} (J^{P}) $ are

\ai
\label{G-6}
a^{hh} (2^{+}) & = & \frac{-1}{4} q^{2} - \frac{1}{4 {\cal K}} \, 
V(\rho) \; , \\
a^{hh} (1^{-}) & = & \frac{1}{4 \, \xi} q^{2} - \frac{1}{4 {\cal K}} \, 
V(\rho) \; , \\
a_{s}^{hh} (0^{+}) & = & \frac{1}{2} q^{2} + \frac{1}{8 {\cal K}} \, 
V(\rho) \; , \\
a_{w}^{hh} (0^{+}) & = & \frac{1}{2 \, \xi} q^{2} - \frac{1}{8 {\cal K}} \, 
V(\rho) \; , \\
a_{sw}^{hh} (0^{+}) & = & \frac{\sqrt{3}}{8 {\cal K}} \, V(\rho) \; = \; 
a_{ws}^{hh} (0^{+}) \; , \\
a_{s1}^{h \fhi} (0^{+}) & = & 
\frac{\sqrt{3}}{4 \, \sqrt{{\cal K}}} \, V'(\rho) \, \fhi^{*}
\; = \; a_{1s}^{\fhi h} (0^{+}) \, , \\ 
a_{s1}^{h \fhi^{*}} (0^{+}) & = & 
\frac{\sqrt{3}}{4 \, \sqrt{{\cal K}}} \, V'(\rho) \, \fhi
\; = \; a_{1s}^{\fhi^{*} h} (0^{+}) \, , \\
a_{w1}^{h \fhi} (0^{+}) & = & \frac{1}{4 \, \sqrt{{\cal K}}} \, 
V'(\rho) \, \fhi^{*}
\; = \; a_{1w}^{\fhi h} (0^{+}) \, , \\  
a_{w1}^{h \fhi^{*}} (0^{+}) & = & \frac{1}{4 \, \sqrt{{\cal K}}} \, 
V'(\rho) \, \fhi
\; = \; a_{1w}^{\fhi^{*} h} (0^{+}) \, , \\ 
a_{s}^{AA} (1^{-}) & = & \sqrt{3} \, ( q^{2} + 2 \, 
e_{o}^{2} \, \rho ) \; ,\\
a_{w}^{AA} (0^{+}) & = & \frac{1}{\al} \, q^{2} + 2 \, 
e_{o}^{2} \, \rho \; ,\\
a_{w1}^{A \fhi} (0^{+}) & = &  e_{o} \, \fhi^{*} \, \qq  \;
=  \; a_{1w}^{\fhi A} (0^{+}) \; , \\
a_{w1}^{A \fhi^{*}} (0^{+}) & = & - e_{o} \, \fhi \, \qq  \;
=  \; a_{1w}^{\fhi^{*} A} (0^{+}) \; , \\
a_{11}^{\fhi\fhi} (0^{+}) & = & V''(\rho) (\fhi^{*})^{2} \; , \\
a_{11}^{\fhi^{*}\fhi} (0^{+}) & = & q^{2} + V'(\rho) + 
\rho V''(\rho) \; = \;
a_{11}^{\fhi\fhi^{*}} (0^{+}) \; , \\
a_{11}^{\fhi^{*}\fhi^{*}} (0^{+}) & = & V''(\rho) \fhi^{2}
\; .
\bj

\vspace{.5cm}

In matrix form, we can write

\be
a(0^{+}) \; = \;
\left( \begin{array}{ccccc}
a_{s}^{hh}(0^{+})         & a_{sw}^{hh}(0^{+})         
& 0                          & a_{s1}^{h \fhi}(0^{+})       
& a_{s1}^{h \fhi^{*}}(0^{+})       \\  
a_{ws}^{hh}(0^{+})        & a_{w}^{hh}(0^{+})          
& 0                          & a_{w1}^{h \fhi}(0^{+})       
& a_{w1}^{h \fhi^{*}}(0^{+})       \\  
0                         & 0                          
& a_{w}^{AA}(0^{+})          & a_{w1}^{A \fhi}(0^{+})       
& a_{w1}^{A \fhi^{*}}(0^{+})       \\  
a_{1s}^{\fhi h}(0^{+})    & a_{1w}^{\fhi h}(0^{+})     
& a_{1w}^{\fhi A}(0^{+})     & a_{11}^{\fhi\fhi}(0^{+})     
& a_{11}^{\fhi\fhi^{*}}(0^{+})     \\  
a_{1s}^{\fhi^{*}h}(0^{+}) & a_{1w}^{\fhi^{*} h}(0^{+}) 
& a_{1w}^{\fhi^{*} A}(0^{+}) & a_{11}^{\fhi^{*}\fhi}(0^{+}) 
& a_{11}^{\fhi^{*}\fhi^{*}}(0^{+})    
       \end{array}
\right) \; , 
\ee

\be
a(1^{-}) \; = \;
\left( \begin{array}{cc}
        a^{hh} (1^{-})  & 0                    \\  
        0               & a_{s}^{AA} (1^{-})   \\  
       \end{array}
\right) \; , 
\ee

\be
a (2^{+}) \; = \; a^{hh} (2^{+})\; .
\ee

\noindent 
To each element $ a^{A B} (J^{^{P}}) $, there is a spin-projector
$ P^{A B} (J^{^{P}}) $. The complete set of projectors for this model
is cast in the appendix. The lower labels $ s, w, $ and $ 1 $ stands for 
transversal, longitudinal and pure-scalar contributions.

Again, the associated average effective potential reads as

\beq
V_{\ka} (\rho) & = & \frac{1}{2} \int \, 
\frac{d^{4} q}{(2 \pi)^{4}} \;
\left[  \; 5 \cdot \, \ln \; \left( \; 
\frac{ a_{\ka} ( 2^{+} ) ( \rho )}{ a_{\ka} 
( 2^{+} ) ( \rho_{\ka} )} \; \right) \,  + \;
 3 \cdot \, \ln \; \left( \; 
\frac{ \det_{\ka} a_{\ka} ( 1^{-} ) ( \rho )}{ 
\det_{\ka} a_{\ka} 
( 1^{-} ) ( \rho_{\ka} )} \; \right)  \, + \,
\right.
\nonumber \\ &   & 
\nonumber \\ & + &
\left. 
\ln \; \left( \; \frac{\det_{\ka} a_{\ka} ( 0^{+} ) 
( \rho )}{\det_{\ka} a_{\ka} 
( 0^{+} ) ( \rho_{\ka} ) } \; \right) \; \right] \; .
\eeq

Given the set of parametrizations (\ref{Parametrizations})
and the average potential above, 
the flow of $ \rho_{\ka} $ and $ \la_{\ka} $ are given by 

\beq
\ga (\ka) &  =  & \frac{-1}{32 \, \pi^{2} \, \ka^{2} \, \la_{\ka} } \;
\int \, d x \, x \;
\ka \, \frac{\partial \, P_{\ka} (x)}{\partial \, \ka} \;
\left[ \; 
5 \; \cdot  \; \frac{\partial}{\partial \, P_{\ka} (x)} \; 
\frac{\partial}{\partial \, \rho} \; \ln \;
\left( \frac{a_{\ka} (2^{+}) (\rho)}{a_{\ka} (2^{+}) (\rho_{\ka})} \; 
\right)_{\rho \, = \, \rho_{\ka}} \; +
\right.
\nonumber \\ &  &
\nonumber \\ & + &
3 \; \cdot  \; \frac{\partial}{\partial \, P_{\ka} (x)} \; 
\frac{\partial}{\partial \, \rho} \; \ln \;
\left( \frac{\det_{\ka} a_{\ka} (1^{-}) (\rho)}{\det_{\ka} 
a_{\ka} (1^{-}) (\rho_{\ka})} \; 
\right)_{\rho \, = \, \rho_{\ka}} \;  +
\\ &  &
\nonumber \\ & + &
\left.
\frac{\partial}{\partial \, P_{\ka} (x)} \; 
\frac{\partial}{\partial \, \rho} \; \ln \;
\left( \frac{\det_{\ka} a_{\ka} (0^{+}) 
(\rho)}{\det_{\ka} a_{\ka} (0^{+}) (\rho_{\ka})}\; 
\right)_{\rho \, = \, \rho_{\ka}} \;
\right] 
\nonumber
\eeq 

\noindent and

\beq
\bet (\ka) &  = & \frac{1}{32 \, \pi^{2}} \;
\int \, d x \, x \;
\ka \, \frac{\partial \, P_{\ka} (x)}{\partial \, \ka} \;
\left[ \; 
5 \; \cdot  \; \frac{\partial}{\partial \, P_{\ka} (x)} \; 
\frac{\partial^{\, 2}}{\partial \, \rho^{\, 2}} \; \ln \;
\left( \frac{a_{\ka} (2^{+}) (\rho)}{a_{\ka} (2^{+}) (\rho_{\ka})} \; 
\right)_{\rho \, = \, \rho_{\ka}} \; +
\right.
\nonumber \\ &  &
\nonumber \\ & + &
3 \; \cdot  \; \frac{\partial}{\partial \, P_{\ka} (x)} \; 
\frac{\partial^{\, 2}}{\partial \, \rho^{\, 2}} \; \ln \;
\left( \frac{\det_{\ka} a_{\ka} (1^{-}) (\rho)}{\det_{\ka} 
a_{\ka} (1^{-}) (\rho_{\ka})} \; 
\right)_{\rho \, = \, \rho_{\ka}} \;  +
\\ &  &
\nonumber \\ & + &
\left.
\frac{\partial}{\partial \, P_{\ka} (x)} \; 
\frac{\partial^{\, 2}}{\partial \, \rho^{\, 2}} \; \ln \;
\left( \frac{\det_{\ka} a_{\ka} (0^{+}) (\rho)}{\det_{\ka} 
a_{\ka} (0^{+}) (\rho_{\ka})}\; 
\right)_{\rho \, = \, \rho_{\ka}} \;
\right] \; .
\nonumber
\eeq 

In the ultraviolet limit, $ \ka^{2} \gg \rho_{\ka} $,
where again $ P_{\ka}(x) \approx \ka^{2} $, we collect the 
terms with highest power of $ P_{\ka}(x) $ and neglect 
$ \rho_{\ka} $ with respect to $ \ka^{2} $.
Thus, we have for the $ \ga $-function

\be
\label{TG-1}
\ka \, \frac{\partial \, \rho_{\ka}}{\partial \, \ka} \; = \; 
\ka^{\, 2} \; \ga (\ka) \; = \; 
\frac{\ka^{2}}{16 \pi^{2} \la_{\ka}} \, \left[ \, 3 e_{o}^{2} 
+ 2 \al e_{o}^{2} + 2 \la_{\ka} \, \right] \, I_{-2}(0) \; .
\ee

\noindent 
This result is the same as in the case without gravity coupling; 
$ \rho_{\ka} $ runs quadratically.

In the infrared limit, $ \ka^{2} \ll \rho_{\ka} $, we get

\be
\label{TG-2}
\ka \, \frac{\partial \, \rho_{\ka}}{\partial \, \ka} \; = \; 
\ka^{\, 2} \; \ga (\ka) \; = \; 
\frac{1}{32 \pi^{2} \la_{\ka}} \, \left[ \,
\frac{3}{2 \rho_{\ka}^{2} e_{o}^{2}} \, I_{\, 0} (0) \,  \ka^{6} +  
\frac{\la_{\ka}}{2} \, I_{\, -2} (0) \, \ka^{2} 
\, \right] \; .
\ee

\noindent 
The first term in the r.h.s. damps
the evolution of $ \rho_{\ka} $ by powers of 
$ \frac{\ka^{\, 2}}{\rho_{\ka}} $.
For $ \rho_{0} \neq 0 $ and $ \la_{0} \neq 0 $, we have 

\be
\label{TG-3}
\rho_{\ka} \; = \; \rho_{0} \left[ \; 1 + \frac{1}{128 \, \pi^{\, 2}} \, 
\frac{I_{0}(0)}{\la_{0} \, e_{o}^{\, 2}} \, 
\frac{\ka^{\, 6}}{\rho_{0}^{\, 3}} \;
\right] \; .
\ee

\noindent 
This exactly agrees with the case without gravity.

Now, the $ \bet $-function, in the ultraviolet limit is given by

\beq
\bet (\ka) &  =  & \frac{1}{32 \, \pi^{2}} \;
\int \, d x \, x \;
\ka \, \frac{\partial \, P_{\ka} (x)}{\partial \, \ka} \;
\left[ \; 
5 \; \cdot \;  
\left( \frac{- \la_{\ka}}{{\cal K} \, P_{\ka}^{\, 2}(x)} \; 
\right) \; +
\right.
\nonumber \\ &  &
\nonumber \\ & + &
3 \; \cdot \;  
\left( \frac{\la_{\ka} \, \xi}{{\cal K} \, P_{\ka}^{\, 2}(x)} + 
\frac{8 \, e_{o}^{\, 4}}{P_{\ka}^{\, 3}(x)} \; 
\right) \;  +
\left(
\frac{\la_{\ka} \, (\xi - 1)}{4 \, {\cal K} \, P_{\ka}^{\, 2}(x)} \; +
\frac{32 \, \al^{\, 2} \, e_{o}^{\, 4}}{P_{\ka}^{\, 3}} \; + 
\right.
\\ &  &
\nonumber \\ & + &
\left.
\left. 
\frac{8 \, \al \, e_{o}^{\, 2} \, \la_{\ka}}{P_{\ka}^{\, 3}} + 
\frac{20 \, \la_{\ka}^{\, 2}}{P_{\ka}^{\, 3}} +
\frac{3 \, \al \, e_{o}^{\, 2} \, \la_{\ka} \, \rho_{\ka} \, (\xi - 1)}
{{\cal K} \, P_{\ka}^{\, 3}} +
\frac{5 \,  \la_{\ka}^{\, 2} \, \rho_{\ka} \, \xi}
{2 \, {\cal K} \, P_{\ka}^{\, 3}}
\right) \;
\right] \; ,
\nonumber
\eeq

\beq
\label{TG-4}
\ka \frac{\partial \, \la_{\ka}}{\partial \, \ka} &  =  &
\frac{1}{32 \, \pi^{2}} \;
\left[ \; 
\left( 
24 \, e_{o}^{\, 4}  + 32 \, \al^{\, 2} \, e_{o}^{\, 4} \, + 
8 \, \al \, e_{o}^{\, 2} \, \la_{\ka} + 20 \, \la_{\ka}^{\, 2}
\right) \; I_{\, - 3}(0)  \; +
\right.
\nonumber \\ &  &
\nonumber \\ & + &
\left( -5 + 3 \, \xi + \frac{( \xi -1 )}{4}
\right) \;  \frac{\la_{\ka} \, \ka^{2}}{{\cal K}} \, I_{\, - 2}(0)  \; +
\\ &  &
\nonumber \\ & + &
\left.
\left(
\frac{3 \, \al \, e_{o}^{\, 2} \, \la_{\ka} \, \rho_{\ka} \, 
( \xi - 1 )}{{\cal K}} \, 
\right)
I_{ - 3}(0) \;  +
\left(
\frac{5 \, \la_{\ka}^{\, 2} \, \rho_{\ka} \, \xi}{2 \, {\cal K}} \, 
\right) \;
I_{\, - 3}(0) \;
\right] \; .
\nonumber
\eeq

\noindent
As $ \ka^{\, 2} \gg \rho_{\ka} $, the last two terms do not
contribute in this limit. Considering $ \ka^{\, 2} \ll {\cal K} $, 
the terms
proportional to $ \frac{\ka^{2}}{{\cal K}} \, I_{\, - 2}(0) $ are not 
relevant, but when the energy scale goes near, 
but below, the Planck's mass,
$ \ka^{\, 2} \approx {\cal K} $, the $ \la_{\ka} $ 
coupling begins to run much 
faster than logarithmically and some care must be taken
as  Einstein theory  at and above  this energy scale needs to be
corrected. 

In the infrared limit,  one gets for the $ \bet $-function

\beq
\bet (\ka) &  =  & \frac{1}{32 \, \pi^{2}} \;
\int \, d x \, x \;
\ka \, \frac{\partial \, P_{\ka} (x)}{\partial \, \ka} \;
\left[ \; 
5 \; \cdot \;  
\left( \frac{- \la_{\ka}}{{\cal K} \, P_{\ka}^{\, 2}(x)} \; 
\right) \; +
\right.
\nonumber \\ &  &
\\ & + &
\left.
3 \; \cdot \;  
\left( \frac{\la_{\ka} \, \xi}{{\cal K} \, P_{\ka}^{\, 2}(x)} + 
\frac{1}{e_{o}^{\, 2} \, \rho_{\ka}^{\, 3}} \; 
\right) \;  +
\left(
\frac{\la_{\ka} \, (\xi + 1)}{2 \, {\cal K} \, P_{\ka}^{\, 2}(x)} \; +
\frac{\la_{\ka}^{\, 2}}{2 \, P_{\ka}^{\, 3}} \; 
\right) \;
\right] \; ,
\nonumber
\eeq

\be
\label{TG-5}
\ka \frac{\partial \, \la_{\ka}}{\partial \, \ka} \;  =  \; 
\frac{1}{32 \, \pi^{2}} \;
\left[ \; 
\frac{(- 9 + 7 \, \xi)}{2} \, \la_{\ka} \,
\frac{\ka^{\, 2}}{{\cal K}} \, I_{\, - 2}(0) \, +  \, 
\frac{3}{e_{o}^{\, 2}} \, \frac{\ka^{\, 6}}{\rho_{\ka}^{\, 3}} \, 
I_{\, 0}(0) \,
+ \, \frac{\la^{\, 2}_{\ka}}{2} \, I_{\, -3}(0) \; 
\right] \; .
\ee

\noindent
The first term in the r.h.s. becomes relevant when the scale is close to 
the Planck's threshold where the Einstein's action is limited to.
The other two terms are the same as those coming from the Scalar-QED model
when gravity corrections are not considered.

\section{Concluding Remarks.}

The first conclusion we can draw from the analysis of these flow 
equations is that the running of the v.e.v. of $ \rho_{\ka} $
is not modified when the gravitational contribution is taken into account,
differently from the $ \la_{\ka} $ case.
Eqs. (\ref{TG-1}), (\ref{TG-2}) and (\ref{TG-3}) are exactly the same 
as eqs. (\ref{sq-6b}), (\ref{sq-9b}) and (\ref{sq-10}), respectively.
We have promoted the metric field $ g_{\mu \nu} $ to a quantum field
without letting the Planck's constant to run; i.e., we took only the
bilinear sector of Einstein's action and defined the effective
action over flat background. In this way, the vacuum structure
of the theory was not modified by gravity-matter couplings; 
the equivalence principle applies. We should mention that in the
context of a higher-order gravity theory, interesting curvature
induced phase transitions can be found between different vacua as
in ref. \cite{Odintsov}.

For the $ \bet $-functions, we can observe that eqs. (\ref{TG-4}) and
(\ref{TG-5}) contain the same terms found in eqs. (\ref{sq-7b}) and
(\ref{sq-12b}) plus corrections when the mass scale is increased.
At $ {\cal K} \gg \ka^{2} \gg \rho_{\ka} $, the flow of $ \la_{\ka} $
is given by the first term in the r.h.s. of eq. (\ref{TG-4}). It is the
same result found in eq. (\ref{sq-7b}). 
For $ {\cal K} \approx \ka^{2} \gg \rho_{\ka} $, the scale is of order 
the Planck's mass and the $ 2^{nd.} $ term of eq. (\ref{TG-4}) becomes
relevant as a correction induced by gravity. 
So, the  $ \bet $-function reads

\beq
\setcounter{equation}{1}
\label{5.1}
\ka \frac{\partial \, \la_{\ka}}{\partial \, \ka} &  =  &
\frac{1}{32 \, \pi^{2}} \;
\left[ \; 
\left( 
24 \, e_{o}^{\, 4}  + 32 \, \al^{\, 2} \, e_{o}^{\, 4} \, + 
8 \, \al \, e_{o}^{\, 2} \, \la_{\ka} + 20 \, \la_{\ka}^{\, 2}
\right) \; I_{\, - 3}(0)  \; +
\right.
\nonumber \\ &  &
\nonumber \\ & + &
\left. 
\left( -5 + 3 \, \xi + \frac{( \xi -1 )}{4}
\right) \;  \frac{\la_{\ka} \, \ka^{2}}{{\cal K}} \, I_{\, - 2}(0)  \; 
\right] \; . 
\eeq

\noindent
The $ 3^{rd.} $ and $ 4^{th.} $
terms of eq. (\ref{TG-4}) would become relevant, compared to
the $ 2^{nd.} $ one, if $ {\cal K} \approx \ka^{2} \approx \rho_{\ka} $ or
$ {\cal K} \approx \rho_{\ka} \gg \ka^{2} $; but these possibilities would
be incompatible with the ultraviolet regime of the $ \bet $-function in the 
limit $ \ka^{2} \gg \rho_{\ka} $.
So, these last two terms can be neglected
and only the terms shown in eq. (\ref{5.1}) enter the game.
If one sticks to gravity as an effective field theory higher-order
terms should be put in, but, as we said before, these do not turn out to
be excited below the Planckian threshold.

On the opposite
range,  $ \rho_{\ka} \gg \ka^{2} $, the infrared one, the first term of
eq. (\ref{TG-5}) appears as a correction by gravity and the last two are
the same found in eq. (\ref{sq-12b}). 
Comparing the behavior of 
the first two terms, one finds two different situations.
If we set up the regime 
\mbox{$ \rho_{\ka}^{3/2} {\cal K}^{-1/2} \ll \ka^{2} \ll \rho_{\ka} $,}
the $ 2^{nd.} $ and $ 3^{rd.} $ terms of eq. (\ref{TG-5}) contribute to 
the \mbox{$ \bet $-function;}
but the more we go into limit $ \ka^{2} \ll \rho_{\ka} $, the $ 2^{nd.} $ one
tends to be neglected at long distances and only the $ 3^{rd.} $ one 
survive the scaling. If 
\mbox{$ \ka^{2} \ll \rho_{\ka}^{3/2} {\cal K}^{-1/2} \ll \rho_{\ka} $,} 
the $ 1^{st.} $
term corrects the case without gravity; but our threshold region is limited
to $ {\cal K} $ as $ \ka^{2} \ll \rho_{\ka} \ll {\cal K} $. So, 
effectively, only the $ 3^{rd.} $ term contributes to the running of 
$ \la_{\ka} $ in this limit. Thus, the behavior of $ \la_{\ka} $ 
in the infrared, with and without gravity, is kept the  same:

\beq
\label{5.2}
\la_{\ka} \, = \, 
\frac{\la_{0}}{ 1 - \frac{\la_{0} I_{-3}(0)}{64 \pi^{2}} \, 
\ln \left( \frac{\ka}{\ka_o} \right) } \; .
\eeq

\noindent
The self-interacting coupling $ \la_{\ka} $ scales slowly to zero
in the deep infrared and the theory still correlates at long
distances as the masses of the gauge particles are suppressed 
with $ \rho_{\ka} $ when $ \ka \rightarrow 0 $ as seen in eq.
(\ref{TG-2}).

Recently, Reuter \cite{Reuter2} found that 
Einstein's gravity antiscreen,
i.e., Newton's constant decreases when $ \ka $ goes to the 
ultraviolet and grows towards 
the infrared with a quadratic running and a negative sign
for it's flow equation as below:

\beq
\ka \frac{\partial \, G_{\ka}}{\partial \, \ka} \, = \,
( - 2 w G_{o} \ka^2 ) \, G_{\ka} \; ,
\eeq

\noindent where $ w $ is a positive number that depends 
on the cutoff functions $ R_{\ka}^{[AB]} $ of eq. (\ref{cutoff})
and $ G_{o} $ is the bare Newton's 
constant. The evolution of the Planck's constant, defined here as
$ {\cal K} = \frac{1}{16 \pi G} $, has an opposite sign to that of $ G $.
So, in eq. (\ref{5.1}), the running of $ \la_{\ka} $ in the ultraviolet
shall behave in the following way: the term of correction by Einstein's
gravity, proportional to $ \frac{\ka^2}{\cal K} $, evolves like
a constant, because $ \cal K $ scales quadratically in the denominator
as does $ \ka^2 $ in the numerator.
Thus, at high-energies, only the first term of eq. (\ref{5.1}) 
shall contribute; analogously to the case without gravity.
$ \la_{\ka} $ runs only logarithmically when $ \cal K $ is shifted 
quadratically to the ultraviolet. The $ \bet $-function in this regime
doesn't feel any correction induced by gravity; at least up to this order
and up to the truncation scheme.

In the infrared, the first term of eq. (\ref{TG-5}) goes down
to a constant due to the same positive quadratic running of
$ \cal K $ and the analysis pointing to eq. (\ref{5.2}) as the
behavior of $ \la_{\ka} $ at low energies is maintained.

Similar results to the  antiscreening of gravity are found 
in ref. \cite{Odintsov2} for different truncations of the space
of actions and some choices of parameters.

We have computed the $ \ga $- and $ \bet $-functions of the 
Scalar-QED corrected by Einstein's gravity. The v.e.v. of
$ \rho_{\ka} $ was found to run quadratically at high
energies and to suppress its running at sufficient low
energies. $ \la_{\ka} $ scales faster than logarithmically 
in the ultraviolet in
the presence of the gravitational coupling 
and logarithmically in the infrared due to the contribution of
massless particles in this limit.

\section*{Acknowledgements}

I would like to thank Prof. Daniele Amati for the kind hospitality
at SISSA/Italy where part of this work was done.
I thank Roberto Percacci for all discussions on this subject
and for sharing with me his insights.
I am grateful to Cl\'ovis Wotzasek and Jos\'e Abdalla 
Helay\"el-Neto for discussions and for reading
the manuscript. This work was supported by the Conselho
Nacional de Desenvolvimento Cient\'\i fico e Tecnol\'ogico/CNPq-Brazil.
The referee is acknowledged for calling attention to ref.\cite{Reuter2}.

\section*{Appendix: }

Below, we list all the spin-projection operators
found in this work.

\beq
P^{hh}(2^{+})_{\mu \nu, \ka \la} & = & \frac{1}{2} \, 
( \, \Th_{\mu \ka} \Theta_{\nu \la} + \Theta_{\mu \la} 
\Theta_{\nu \ka} ) - 
\frac{1}{3} \Theta_{\mu \nu} \Theta_{\ka \la}
\; , \nonumber \\
P^{hh}(1^{-})_{\mu \nu, \ka \la} & = & 
\frac{1}{2} ( \Th_{\mu \ka} \om_{\nu \la} + \Th_{\mu \la} \om_{\nu \ka} + 
\Th_{\nu \ka} \om_{\mu \la} + \Th_{\nu \la} \om_{\mu \ka} )
 \; , \nonumber \\
P^{hh}_{s}(0^{+})_{\mu \nu, \ka \la} & = & 
\frac{1}{3} \Th_{\mu \nu} \Th_{\ka \la}
\; , \nonumber \\
P^{hh}_{w}(0^{+})_{\mu \nu, \ka \la} & = & 
\om_{\mu \nu} \om_{\ka \la}
\; , \nonumber \\
P^{hh}_{sw}(0^{+})_{\mu \nu, \ka \la} & = & 
\frac{1}{\sqrt{3}} \Th_{\mu \nu} \om_{\ka \la}
\; , \nonumber \\  
P^{hh}_{ws}(0^{+})_{\mu \nu, \ka \la} & = &
\frac{1}{\sqrt{3}} \om_{\mu \nu} \Th_{\ka \la}
\; , \nonumber \\
P_{s1}^{h \fhi}(0^{+})^{\mu \nu} & = & 
\frac{1}{\sqrt{3}} \, \Th^{\mu \nu}
\; , \nonumber \\
P_{1s}^{\fhi h}(0^{+})^{\mu \nu} & = & 
\frac{1}{\sqrt{3}} \, \Th^{\mu \nu}
\, , \nonumber \\ 
P_{s1}^{h \fhi^{*}}(0^{+})^{\mu \nu} & = & 
\frac{1}{\sqrt{3}} \, \Th^{\mu \nu}
\, , \nonumber \\ 
P_{1s}^{\fhi^{*} h}(0^{+})^{\mu \nu} & = & 
\frac{1}{\sqrt{3}} \, \Th^{\mu \nu}
\, , \nonumber \\ 
P_{w1}^{h \fhi}(0^{+})^{\mu \nu} & = & 
\om^{\mu \nu}
\; , \nonumber \\
P_{1w}^{\fhi h}(0^{+})^{\mu \nu} & = & 
\om^{\mu \nu}
\, , \nonumber \\ 
P_{w1}^{h \fhi^{*}}(0^{+})^{\mu \nu} & = & 
\om^{\mu \nu}
\; , \nonumber \\
P_{1w}^{\fhi^{*} h}(0^{+})^{\mu \nu} & = &
\om^{\mu \nu}
\; , \nonumber \\
P_{s}^{AA}(1^{-})^{\mu \nu} & = & \frac{1}{\sqrt{3}} \, \Th^{\mu \nu}
\; , \nonumber \\
P_{w}^{AA}(0^{+})^{\mu \nu} & = & \om^{\mu \nu}
\; , \nonumber \\
P_{w1}^{A \fhi}(0^{+})^{\mu \, .} & = & 
\om^{\mu \nu} \, \mbox{\^q}_{\nu}
\; , \nonumber \\
P_{1w}^{\fhi A}(0^{+})^{\mu \, .} & = & 
\om^{\mu \nu} \, \mbox{\^q}_{\nu}
\; , \nonumber \\
P_{w1}^{A \fhi^{*}}(0^{+})^{\mu \, .} & = &  
\om^{\mu \nu} \, \mbox{\^q}_{\nu}
\; , \nonumber \\
P_{1w}^{\fhi^{*} A}(0^{+})^{\mu \, .} & = &  
\om^{\mu \nu} \, \mbox{\^q}_{\nu}
\; , \nonumber \\
P_{11}^{\fhi\fhi}(0^{+}) & = & 
1
\; , \nonumber \\
P_{11}^{\fhi\fhi^{*}}(0^{+}) & = &
1
\; , \nonumber \\
P_{11}^{\fhi^{*}\fhi}(0^{+}) & = &
1
\; , \nonumber \\
P_{11}^{\fhi^{*}\fhi^{*}}(0^{+}) & = & 
1
\; \nonumber  .
\eeq

\noindent
The operators $ \Th^{\mu \nu} $ and $ \om^{\mu \nu} $ stand for 
the usual transverse and longitudinal projectors on the space of
vectors, 

\beq
\mbox{\^q}^{\mu} = 
\frac{\mbox{q}^{\mu}}{\sqrt{\mbox{q}^{2}}} \; , \:\:\:
\de^{\mu \nu} = \Th^{\mu \nu} + \om^{\mu \nu} \:\:\: \mbox{and} \:\:\: 
\om^{\mu \nu} = \mbox{\^q}^{\mu} \mbox{\^q}^{\nu} \, , \nonumber
\eeq

\noindent and are identified by the lower labels $ s $ and $ w $, 
as in the Barnes and Rivers notation \cite{Barnes}.
The lower index $ 1 $, label the unity contribution from the scalar 
fields to the spin operators.

\end{document}